\documentclass[preprintnumbers]{revtex4}

\usepackage{graphicx}
\usepackage{epsf}

\usepackage{amsmath,amssymb}
\newcommand{\bvec}{\boldsymbol}

\begin{document}
\preprint{KUNS-2562}
\title{$^{10}$B+$\alpha$ states with chain-like structures in $^{14}$N}
\author{Yoshiko Kanada-En'yo}
\affiliation{Department of Physics, Kyoto University, Kyoto 606-8502, Japan}

\begin{abstract}
I  investigate $^{10}$B+$\alpha$ cluster states of $^{14}$N with a $^{10}$B+$\alpha$ 
cluster model. Near the $\alpha$-decay threshold energy,
I obtain $K^\pi=3^+$ and $K^\pi=1^+$ rotational bands
having $^{10}$B($3^+$)+$\alpha$ and $^{10}$B($1^+$)+$\alpha$ components, respectively.
I assign the band-head state of the $K^\pi=3^+$ band to the 
experimental $3^+$ at $E_x$=13.19 MeV of $^{14}$N 
observed in $\alpha$ scattering reactions by $^{10}$B and show that the calculated $\alpha$-decay
width is consistent with the experimental data.
I discuss an $\alpha$-cluster motion around the $^{10}$B cluster and show that 
$^{10}$B+$\alpha$ cluster states contain significant components of a linear-chain 3$\alpha$ configuration, 
in which an $\alpha$ cluster is localized in the longitudinal direction around the deformed $^{10}$B cluster.
\end{abstract}

\maketitle
\section{Introduction}
%Cluster is one of the essential features of nuclei.
It is known that cluster structures appear in various nuclei including unstable nuclei
(for instance, \cite{Oertzen-rev,AMDsupp,AMDrev1,AMDrev2,Horiuchi-rev} and references therein.) 
For cluster states having an $\alpha$ cluster around a core nucleus, well-known 
examples are $^{16}$O+$\alpha$ states in $^{20}$Ne and $^{12}$C+$\alpha$ states in $^{16}$O \cite{Fujiwara-supp}.
Recent experimental and theoretical studies have revealed many cluster resonances
in highly excited states near the $\alpha$-decay threshold also in unstable 
nuclei, for instance, $^{A-4}$He+$\alpha$ states in Be isotopes 
\cite{Oertzen-rev,AMDrev2,SEYA,OERTZEN,ARAI,Dote:1997zz,ENYObe10,ITAGAKI,OGAWA,KanadaEn'yo:2002ay,Descouvemont02,Ito:2003px,Ito:2005yy,FREER,FREER-2,SAITO04,Curtis:2004wr,Millin05,Freer:2006zz,Bohlen:2007qx,Curtis:2009zz,Yang14},
$^{10}$Be+$\alpha$ states in $^{14}$C \cite{Soic:2003yg,oertzen04,Price:2007mm,Haigh:2008zz,Suhara:2010ww}, 
$^{14}$C+$\alpha$  states in $^{18}$O and their mirror states 
\cite{Gai:1983zz,Descouvemont:1985zz,Gai:1987zz,Curtis:2002mg,Ashwood:2006sb,Yildiz:2006xc,Furutachi:2007vz,Fu:2008zzf,Johnson:2009kj,oertzen-o18}, 
$^{18}$O+$\alpha$ states in $^{22}$Ne 
\cite{Scholz:1972zz,Descouvemont:1988zz,Rogachev:2001ti,Goldberg:2004yk,Curtis:2002mg,Ashwood:2006sb,Yildiz:2006xc,Kimura:2007kz}.

Multi-$\alpha$ cluster states such as cluster gas and linear-chain states of $n\alpha$ systems 
are also interesting topics. The $\alpha$-cluster gas
was  proposed by Tohsaki {\it et al.} to describe $3\alpha$ cluster structure of $^{12}$C($0^+_2$)
\cite{Tohsaki:2001an} and extended to excited states of $^{12}$C and 
other nuclei \cite{Funaki:2003af,Yamada:2003cz,Funaki:2008gb}.
The linear-chain $n\alpha$ state was originally proposed for 
$^{12}$C($0^+_2$)  by Morinaga in the 1950-60s \cite{morinaga56,morinaga66}. However, in the 1970s,
this picture was excluded at least for $^{12}$C($0^+_2$) having a larger $\alpha$-decay width
than the one expected from the linear-chain structure \cite{suzuki72}.
In spite of many discussions for several decades, existence of linear-chain $n\alpha$ states have yet been confirmed 
and it is still an open problem to be solved.
It is naively expected that the linear-chain configuration is not favored in an 
$n\alpha$ system because it costs much kinetic energy to keep $\alpha$
clusters in a row. It means that some mechanism is necessary to form the linear-chain structure.
In progress of physics of unstable nuclei since the 1990s,
it was proposed for neutron-rich C isotopes  that 
excess neutrons may stabilize the linear-chain structure \cite{Oertzen-rev,OERTZEN}.
%Z. Phys. A 357, 355-365 (1997).  Oertzen-rev
Itagaki {\it et al.} analyzed stability of a $3\alpha$-chain configuration surrounded by excess neutrons in molecular orbitals
against the bending motion and
suggested that the linear-chain structure can be stable in $^{16}$C but
unstable in $^{12}$C and $^{14}$C \cite{Itagaki:2001mb}. 
More recently, Suhara and the author predicted 
a rotational band with a linear $3\alpha$ chain configuration in excited states of $^{14}$C
near the $\alpha$-decay threshold \cite{Suhara:2010ww}. They pointed out that 
the orthogonal condition to lower states is important for the stability of the linear-chain structure.
The linear-chain structure is expected to be more favored in high spin states 
because of stretching effect in rotating systems as suggested in $^{15}$C \cite{Oertzen-rev}
and $^{16}$O \cite{Ichikawa:2011iz}.

According to analysis in Refs.~\cite{Suhara:2010ww,Suhara:2011cc}, 
linear-chain states of $^{14}$C are found to have a $2\alpha+2n$ correlation and are interpreted as 
$^{10}$Be+$\alpha$ structures, where the $^{10}$Be cluster is a prolately deformed state 
containing a 2$\alpha$ core and an additional $\alpha$ cluster is located in the longitudinal direction of the $^{10}$B cluster.
Similarly, the linear-chain state of $^{15}$C suggested in Ref.~\cite{Oertzen-rev} also shows 
a $^{11}$Be+$\alpha$ cluster structure with a prolately deformed 
$^{11}$Be cluster and an $\alpha$ cluster in the longitudinal direction.
It means that, the linear-chain states in these neutron-rich C tend to have the $2\alpha$ correlation, and therefore 
$3\alpha$ linear-chain structures are expected to be found in Be+$\alpha$ cluster states.

In this paper, I focus on $^{10}$B+$\alpha$ cluster states in excited states of $^{14}$N.
In experimental energy levels of $^{14}$N near the $\alpha$-decay threshold, $J^\pi=3^+$ and $1^+$ resonances
were observed by $\alpha$ elastic scattering by $^{10}$B \cite{Mo:1973zzb}. 
These resonances are expected to be 
$^{10}$B+$\alpha$ cluster states because of significant $\alpha$-decay widths.  
In analogy to $^{10}$Be+$\alpha$ cluster states, it is interesting to investigate whether 
$^{10}$B+$\alpha$ cluster states with the dominant linear-chain structure exist.
The ground state ($3^+$) and the first excited state ($1^+$) of $^{10}$B can be 
described by the deformed state with a 2$\alpha$ core surrounded by $pn$ 
as discussed in  Refs.~\cite{SEYA,Kanada-En'yo:2015cia}. 
If a $^{10}$B+$\alpha$ cluster state has an $\alpha$ cluster in the longitudinal direction of
the deformed $^{10}$B cluster, the $^{10}$B+$\alpha$ cluster state can be interpreted as a kind of 
linear-chain state that contains dominantly $3\alpha$ clusters arranged in a row.

My aim is to study $^{10}$B+$\alpha$ cluster states of $^{14}$N
and discuss $3\alpha$ configurations, in particular, the linear-chain component in 
 $^{10}$B+$\alpha$ cluster states. 
I calculate $^{10}{\rm B}(3^+)\otimes L_\alpha$ and $^{10}{\rm B}(1+)\otimes L_\alpha$ components and 
evaluate partial $\alpha$-decay widths of $^{10}$B+$\alpha$ cluster states.
To discuss stability of the linear-chain $^{10}$B+$\alpha$ structure, 
I analyze angular motion of an $\alpha$ cluster around the deformed $^{10}$B cluster, i.e., 
rotation of the $^{10}$B cluster. 

This paper is organized as follows. In Sec.~\ref{sec:formulation}, 
I explain the formulation of the present $^{10}$B+$\alpha$ 
cluster model. In Sec.~\ref{sec:results},  calculated positive-parity states and $E2$ transition strengths of $^{14}$N
are shown. 
I discuss $\alpha$ cluster motion around $^{10}$B($3^+$) and $^{10}$B($1^+$) 
in Sec.~\ref{sec:discussion}. Finally, a summary is given in Sec.~\ref{sec:summary}.

\section{Formulation of  the $^{10}$B+$\alpha$ cluster model}\label{sec:formulation}
%We apply the $^{10}$B+$\alpha$ cluster model to investigate 
% $^{10}$B+$\alpha$ cluster states in $^{14}$N.

\subsection{Description of the $^{10}$B cluster} 
For the $^{10}$B cluster in the present $^{10}$B+$\alpha$ cluster model, 
I adopt a $2\alpha+(pn)$ wave function 
which can reasonably describe features of the ground ($J^\pi=3^+$) and first excited ($1^+$) states of $^{10}$B
as discussed in Ref.~\cite{Kanada-En'yo:2015cia}. The $2\alpha+(pn)$ wave function is 
given by a three-body cluster wave function, where $\alpha$ clusters and a dinucleon 
$(pn)$ cluster are written by $(0s)^4$ and $(0s)^2$ harmonic oscillator configurations, respectively, as
\begin{eqnarray}
\Phi_{2\alpha+pn}(\bvec{R}_1,\bvec{R}_2,\bvec{R}_3)&=&
{\cal A}\lbrace\Phi_\alpha(\bvec{R}_1)\Phi_\alpha(\bvec{R}_2)\Phi_{pn}(\bvec{R}_3)\rbrace,\\
\Phi_\alpha(\bvec{R})&=&\psi_{p\uparrow}(\bvec{R})\psi_{p\downarrow}(\bvec{R})
\psi_{n\uparrow}(\bvec{R})\psi_{n\downarrow}(\bvec{R}),\\
\Phi_{pn}(\bvec{R})&=&\psi_{p\uparrow}(\bvec{R})\psi_{n\uparrow}(\bvec{R}),\\
\psi_{\sigma}(\bvec{R})&=&\varphi_{0s}(\bvec{R})\chi_{\sigma},
\end{eqnarray}
where ${\cal A}$ is the antisymmetrizer for all nucleons and $\varphi_{0s}(\bvec{R})$ is the spatial part of the single-particle 
wave function of the $0s$ orbit around $\bvec{R}$;
\begin{equation}
\varphi_{0s}(\bvec{R})=
\left(\frac{2\nu}{\pi}\right)^{3/4}\exp\left\{-\nu(\bvec{r}-\bvec{R})^2\right\},
\end{equation}
and $\chi_{\sigma}$ is the spin-isospin wave function for 
$\sigma=p\uparrow$, $p\downarrow$, $n\uparrow$, and $n\downarrow$.
For the $^{10}$B cluster,  I set 2 $\alpha$ clusters in the $z$ direction 
as $\bvec{R}_1-\bvec{R}_2=(0,0,d_{2\alpha})$ with $d_{2\alpha}=3$ fm 
and a spin-aligned $pn$ cluster on the $x$-$y$ plane at the distance $d$ 
from the $2\alpha$ center as $\bvec{R}_3-(\bvec{R}_1+\bvec{R}_2)/2=(d\cos\phi,d\sin\phi,0)$.
I write the $^{10}$B wave function localized 
around $\bvec{X}_B\equiv (4\bvec{R}_1+4\bvec{R}_2+2\bvec{R}_3)/10$
as $\Phi_{^{10}{\rm B}}(\bvec{X}_B; d,\phi)$ with the center position $\bvec{X}_B$
and the distance and angle parameters, $d$ and $\phi$, for the $pn$ cluster position.
In the $^{10}$B+$\alpha$ cluster
model, I superpose the $^{10}$B wave  functions with $d=1,2$ (fm) and 
$\phi_j=\frac{\pi}{4}(j-0.5)$ ($j=1,\ldots,8$).
%\begin{equation}
%\sum_{i=1,2} \sum_{j=1,\ldots,8} C_{i,j} \Phi_{^{10}{\rm B}}
%(d_i,\phi_j).
%\end{equation}
Parity ($\pi$) and $K$ ($I_z$) projections of the subsystem $^{10}$B can be
approximately done by the  $\phi_j$ summation;
\begin{equation} 
\Phi_{^{10}{\rm B}({I^\pi_z})}(\bvec{X}_B;d)=\sum_j c_j  \Phi_{^{10}{\rm B}}
(\bvec{X}_B;d,\phi_j)
\end{equation}
with $c_j=\exp(i(I_z-1)\phi_j)$ and $\pi=(-1)^{I_z-1}$. Here, 
$I_z$ is the $z$ component of the total angular momentum $\bvec{I}$ of
$^{10}$B and is given by a sum of the aligned intrinsic 
spin $S_z=1$ and the orbital $\phi$ rotation of the $pn$ cluster.
It is clear that $\phi_j$ superposition with given coefficients 
$c_j$ is equivalent to $I_z$ mixing.

\subsection{$^{14}$N wave function in the $^{10}{\rm B}+\alpha$ model}
\begin{figure}[htb]
\begin{center}
\includegraphics[width=7cm]{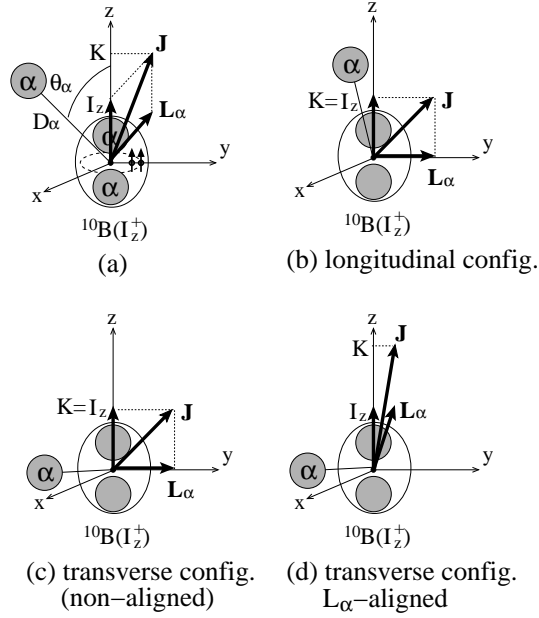} 	
\end{center}
%\vspace{0.5cm}
  \caption{\label{fig:10B-4He}
Schematic figures for $^{10}$B+$\alpha$ configurations. (a)  Parameters of the $^{10}$B+$\alpha$ cluster model.
(b) Longitudinal configuration for $\theta_\alpha\sim 0$, (c) transverse configuration for $\theta_\alpha\sim
\pi/2$ without $L_\alpha$ alignment, and (d) $L_\alpha$-aligned transverse configuration. 
}
\end{figure}

A $^{10}{\rm B}+\alpha$ wave function is written using
the $^{10}$B wave function $\Phi_{^{10}{\rm B}}(\bvec{X}_B;d,\phi)$ 
and the $\alpha$-cluster wave function $\Phi_\alpha(\bvec{X}_\alpha)$
as
\begin{equation}
\Phi_{^{10}{\rm B}+\alpha}(D_\alpha,\theta_\alpha; d,\phi) =
{\cal A} \left\lbrace\Phi_{^{10}{\rm B}}(\bvec{X}_B;d,\phi)   
\Phi_\alpha(\bvec{X}_\alpha) \right\rbrace,
\end{equation}
where $\bvec{X}_\alpha-\bvec{X}_B=(D_\alpha \sin\theta_\alpha, 0, D_\alpha \cos\theta_\alpha)$.
Here the distance $D_\alpha$ and the angle $\theta_\alpha$ indicate 
the $\alpha$-cluster position relative to the deformed $^{10}{\rm B}$ cluster (see Fig.~\ref{fig:10B-4He}).  
The center of mass position  
is taken to be $4\bvec{X}_\alpha+10\bvec{X}_B=0$ so as to decouple 
the center of mass motion and the intrinsic wave function.  Wave functions for the
$J^\pi_n$ states of $^{14}$N are expressed by superposition of the $J^\pi$-projected 
wave functions as 
\begin{equation}\label{eq:14N}
\Psi_{^{14}\textrm{N}(J^\pi_n)}=\sum_K \sum_{D_\alpha,\theta_\alpha} \sum_{d,\phi} 
C(K,D_\alpha,\theta_\alpha,d,\phi) 
\hat P^{J\pi}_{MK} \Phi_{^{10}{\rm B}+\alpha}(D_\alpha,\theta_\alpha; d,\phi), 
\end{equation}
where $\hat P^{J\pi}_{MK}$ is the parity and total angular momentum projection 
operator. Coefficients $C(K,D_\alpha,\theta_\alpha,d,\phi)$ are determined
by diagonalizing Hamiltonian and norm matrices.
I take $D_\alpha=\{2,\ldots, 6\}$ (fm), $\theta_\alpha=\{0, \pi/4, \pi/2\}$,  $d=\{1,2\}$ (fm), and
$\phi=\frac{\pi}{4}(j-0.5)$ ($j=1,\ldots,8$). In the present paper, I calculate positive-parity 
($\pi=+$) states of $^{14}$N.

In Eq.~\eqref{eq:14N}, coupling of 
$\bvec{I}$ (the spin of the $^{10}$B cluster) and 
$\bvec{L}_\alpha$ (the orbital angular momentum of the $\alpha$ cluster relative to the $^{10}$B cluster)
is implicitly described by the $J^\pi$ projection, $K$ mixing,
and  $\theta_\alpha$, $\phi$ summations.
As shown in Fig.~\ref{fig:10B-4He},  
$\bvec{L}_\alpha$ couples with $\bvec{I}$ to the total angular momentum
$\bvec{J}=\bvec{L}_\alpha+\bvec{I}$. The $z$ component, $J_z=I_z+L_{\alpha z}$, 
is the so-called $K$ quantum.
Strictly speaking, 
$L_\alpha=0,2$ ($S,D$-wave) mixing 
is approximately taken into account by the summation of $\theta_\alpha=\{0, \pi/4, \pi/2\}$
but higher $L_\alpha(\ge 4)$ mixing can not be controlled in the present calculation
because of the finite number of mesh points for $\theta_\alpha$. 

%%%%%%%%%%%%%%%%%%%%%%%%%%%%%%
%%%%%%%%%%%%%%%%%%%%%%%%%%%%%

\subsection{Overlap function and $\alpha$-cluster probability}
In order to evaluate $^{10}$B$(3^+)\otimes L_\alpha$ and $^{10}$B$(1^+)\otimes L_\alpha$
components at a channel radius ($D_\alpha$),
I consider the
$L_\alpha L_{\alpha z}$ projected 
$^{10}{\rm B}(I^\pi_z)+\alpha$ wave function, 
\begin{eqnarray}\label{eq:Lalpha-projection}
| J^\pi K;^{10}{\rm B}(I^\pi_z);D_\alpha,L_\alpha L_{\alpha z}\rangle 
&=&n_0\sum_{\theta_\alpha}\omega(\theta_\alpha) 
Y^{L_\alpha}_{L_{\alpha z}}(\theta_\alpha) \hat P^{J\pi}_{MK}
\Phi_{^{10}{\rm B}(I^\pi_z)+\alpha}(D_\alpha,\theta_a),\\
\Phi_{^{10}{\rm B}(I^\pi_z)+\alpha}(D_\alpha,\theta_a) 
&=&{\cal A}\left\lbrace \Phi_{^{10}{\rm B}(I^\pi_z)}(\bvec{X}_B;d) 
\Phi_\alpha(\bvec{X}_\alpha) \right\rbrace,
%\hat P_\alpha(L_\alpha L_{\alpha z}) 
%\hat P^{J+}_{MK} {\cal A} \left\lbrace\Phi_{^{10}{\rm B}(I_z}(\bvec{X}_B;d)   
%\Phi_\alpha(\bvec{X}_\alpha) \right\rbrace
\end{eqnarray}
with $I_z=\{1,3\}$, $\pi=+$, $K=I_z+L_{\alpha z}$, 
$\bvec{X}_\alpha-\bvec{X}_B=(D_\alpha \sin\theta_\alpha, 0, D_\alpha \cos\theta_\alpha)$,
and $4\bvec{X}_\alpha+10\bvec{X}_B=0$. 
$Y^\lambda_\mu$ is the 
spherical harmonics.
$\Phi_{^{10}{\rm B}(I^\pi_z)+\alpha}(D_\alpha,\theta_a)$ is the wave function for the $\alpha$ cluster
at $(D_\alpha,\theta_\alpha)$ around the $I_z$ projected $^{10}$B cluster, for which
I fix $d=2$ fm in the present analysis. $n_0$ is
determined from the normalization condition $\langle JK;^{10}{\rm B}(I^\pi_z);D_\alpha,L_\alpha L_{\alpha z}| JK;^{10}{\rm B}(I^\pi_z);D_\alpha,L_\alpha L_{\alpha z}\rangle=1$.
In Eq.~\eqref{eq:Lalpha-projection}, the $L_\alpha L_{\alpha z}$
projection is approximately performed 
by the summation $\theta_\alpha=\frac{\pi}{N_\theta}i$ $(i=0,\ldots,N_\theta)$ with the    
weight function 
$\omega(\theta_\alpha)=\int^{\max[\theta_\alpha+\pi/2N_\theta,\pi]}_{\min[\theta_\alpha-\pi/2N_\theta,0]}\sin{\theta}d\theta$. 
I perform only $L_\alpha=0, 2 $ projections because $L_\alpha\ge 4$ projections are not possible 
for the present $N_\theta=4$ case.
I calculate the squared overlap of the $^{14}$N wave function with the above
wave function,
$|\langle JK;^{10}{\rm B}(I^\pi_z);D_\alpha,L_\alpha L_{\alpha z} | \Psi_{^{14}{\rm N}(J^\pi_n)}\rangle|^2$. Assuming that the $3^+_1$ and $1^+_1$ states of the $^{10}$B cluster are approximately described by 
the $I_z$ projected $^{10}$B wave functions, $^{10}$B($I^\pi_z=3^+$) and  $^{10}$B($I^\pi_z=1^+$),
respectively, I approximately estimate the $^{10}{\rm B}(I^\pi)\otimes (L_\alpha=0,2) $ components 
as 
\begin{equation}\label{eq:Lalpha}
P_{^{10}{\rm B}(I^\pi)\otimes L_\alpha}(D_\alpha)\approx\sum_{L_{\alpha z}} | \langle JK | II_zL_\alpha L_{\alpha z}\rangle  
\langle JK;^{10}{\rm B}(I_z);D_\alpha,L_\alpha L_{\alpha z}| \Psi_{^{14}{\rm N}(J^\pi_n)}\rangle |^2
\end{equation}
with $I_z=I$, where  $\langle JK | II_zL_\alpha L_{\alpha z}\rangle$
is the Clebsch-Gordan coefficient.

%Instead of the $L_\alpha$ projection described above,
I also calculate $\alpha$-cluster probability at $(D_\alpha,\theta_\alpha)$
around the $I_z$ projected $^{10}$B cluster
as 
\begin{eqnarray}
\label{eq:P_theta}
P(JK; ^{10}{\rm B}(I^\pi_z); D_\alpha, \theta_\alpha)&=&|\langle JK;D_\alpha,\theta_\alpha;
^{10}{\rm B}(I^\pi_z) | \Psi_{^{14}{\rm N}(J^\pi_n)}\rangle|^2,\\
| JK;^{10}{\rm B}(I^\pi_z);D_\alpha,\theta_\alpha\rangle 
&=&n_0 \hat P^{J+}_{MK}\Phi_{^{10}{\rm B}(I^\pi_z)+\alpha}(D_\alpha,\theta_a) .
\end{eqnarray}
The probability $P(JK; ^{10}{\rm B}(I^\pi_z); D_\alpha, \theta_\alpha)$ is useful 
to discuss geometric configurations of $3\alpha$ clusters in $^{10}$B+$\alpha$ cluster states 
in the strong coupling picture. For instance, $P(JK; ^{10}{\rm B}(I^\pi_z); D_\alpha, \theta_\alpha)$
for $\theta_\alpha\sim 0$ means the component of  the ``longitudinal'' configuration, where 
the $\alpha$ cluster is localized in the longitudinal direction of the deformed $^{10}{\rm B}(I^\pi_z)$ cluster.
This configuration corresponds to the linear-chain structure as 3 $\alpha$ clusters are 
arranged in a row as shown in Fig.~\ref{fig:10B-4He}(b). Because of the axial symmetry, the longitudinal configuration contains only $K=I_z$ $(L_{\alpha z}=0)$ component. 
For $\theta_\alpha \sim \pi/2$, $P(JK; ^{10}{\rm B}(I^\pi_z); D_\alpha, \theta_\alpha)$
indicates the component of the ``transverse configuration'' for the $\alpha$ cluster 
in the transverse direction of the deformed $^{10}{\rm B}(I^\pi_z)$ cluster. The transverse configuration
contains $K\ne I_z$ components corresponding to the 
alignment of $\bvec{L}_\alpha$ to the spin of the $pn$-cluster ($I_z$) in the $^{10}{\rm B}$ cluster
as well as the $K=I_z$ component (see Fig.~\ref{fig:10B-4He}(b) and (c)).

\section{Results}\label{sec:results}

%%%%%%%%%%%%%%%%%%%%%%%%%%%%%%
\begin{figure}[htb]
\begin{center}
\includegraphics[width=7.cm]{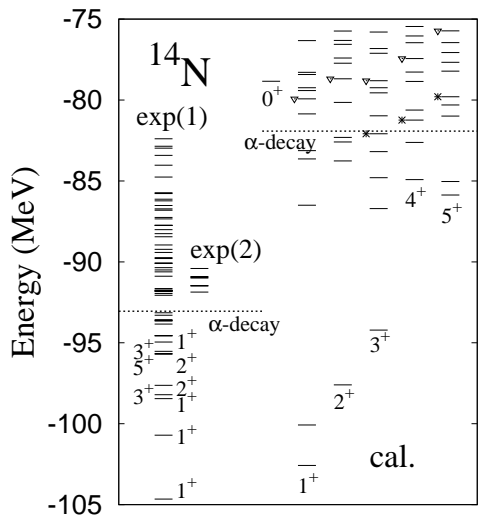} 	
\end{center}
%\vspace{0.5cm}
  \caption{\label{fig:spe} Positive-parity energy levels of $^{14}$N obtained by
the $^{10}$B+$\alpha$ cluster model compared with experimental levels taken from \cite{AjzenbergSelove:1991zz}. 
$^{10}$B+$\alpha$ cluster states in the $K^\pi=3^+$ band and those in the $K^\pi=1^+$ band
are labeled by asterisk and down-triangle symbols, respectively.
}
\end{figure}
%%%%%%%%%%%%%%%%%%%%%%%%%%%%%

%%%%%%%%%%%%%%%%%%%%%%%%%%%%%%
\begin{figure}[htb]
\begin{center}
\includegraphics[width=7.0cm]{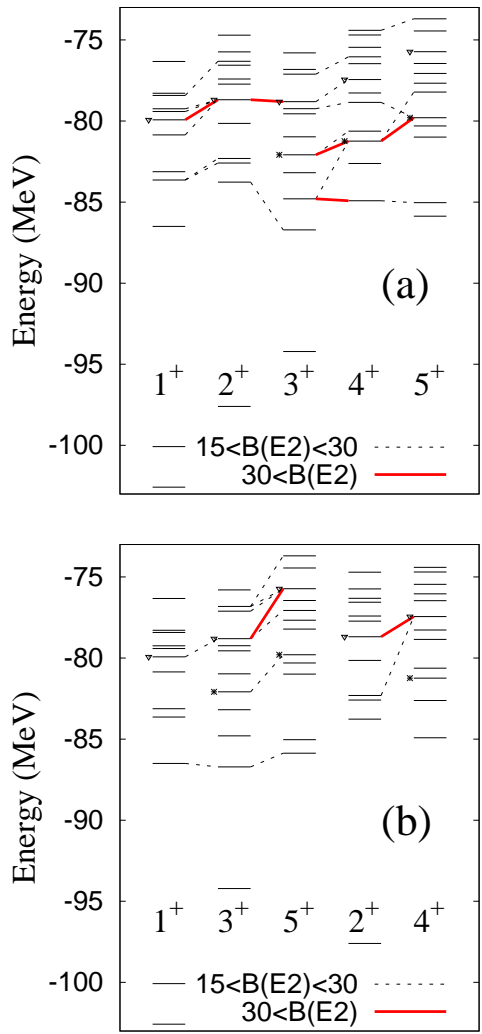} 	
\end{center}
%\vspace{0.5cm}
  \caption{(Color online)
 $E2$ transition strengths calculated by the $^{10}$B+$\alpha$ cluster model 
for (a) $J^+\rightarrow J^+-1$ and (b) $J^+\rightarrow J^+-2$ transitions with 
$B(E2)\ge 15 \textrm{e}^2{\rm fm}^4$.
Asterisk and down-triangle symbols show $^{10}$B+$\alpha$ cluster states
 in the $K^\pi=3^+$ and $K^\pi=1^+$ bands, respectively. 
\label{fig:be2}}
\end{figure}
%%%%%%%%%%%%%%%%%%%%%%%%%%%%%

I adopt the two-body effective nuclear interactions used in Ref.~\cite{Kanada-En'yo:2015cia} which are adjusted to 
describe low-lying energy levels of $^{10}$B. Namely, I use the Volkov central force
\cite{VOLKOV}
with the Bartlett, Heisenberg, and Majorana parameters $b=h=0.06$ and $m=0.60$
and the G3RS spin-orbit force \cite{LS} with the strength $u_{I}=-u_{II}=1300$ MeV,
and the Coulomb force approximated by seven Gaussians.
Using these interactions, Energies of  $^{10}$B are obtained 
to be $-$54.3 MeV for the ground state ($3^+$) and $-53.4$ MeV for the first excite state ($1^+$)
with the $2\alpha+pn$ cluster model by superposing 
$\sum_{I_z,d}\hat P^{I\pi}_{MI_z}\Phi_{^{10}{\rm B}}(\bvec{X}_B=0;d,\phi=0)$ with $d=1,2$ (fm).
Though the calculation underestimates the experimental binding energy (64.75 MeV),
it reproduces the spin parity  of the ground state ($^{10}$B$(3^+_{\rm g.s.})$), and 
also the calculated excitation energy $E_x=0.9$ MeV of the 
$1^+$ state reasonably agrees to the experimental value $E_x=0.72 $ MeV for $^{10}$B$(1^+_1)$.

Using the $^{10}$B+$\alpha$ cluster wave function in Eq.~\eqref{eq:14N},
I calculate positive-parity states of $^{14}$N. Properties of the  
ground state $^{14}$N$(1^+_{\rm g.s.})$ are reasonably reproduced by the present calculation.
Namely, the binding energy B.E.=102.6 MeV, the magnetic moment $\mu=0.36$ ($\mu_N$), and the 
electric quadrupole moment $Q=2.4$ ($e{\rm fm}^2$) of $^{14}$N($1^+_{\rm g.s.}$), 
obtained by the present calculation reasonably agree to the experimental data 
(B.E.=104.66 MeV, $\mu=0.4038$ ($\mu_N$), and $Q=1.93(8)$  ($e{\rm fm}^2$)).
The calculated energy spectra are shown in Fig.~\ref{fig:spe}. The $\alpha$-decay threshold is much 
higher in the present calculation than the experimental threshold. 
In other words, the ground and some low-lying states of $^{14}$N show
too deep binding from the $\alpha$-decay threshold compared with the experimental data.
The significant overestimation of the $\alpha$-decay threshold is a general problem
in microscopic calculations with density-independent two-body effective interactions
as found for $^{14}$C and O isotopes \cite{Fujiwara-supp,Descouvemont:1985zz,Suhara:2010ww}. 
One of the origins of this problem is a difficulty in reproducing systematics of binding energies in a wide mass-number region with such effective interactions.
In this paper, I mainly investigate $^{10}{\rm B}+\alpha$ cluster states 
near the $\alpha$-decay threshold and discuss their features.
In the calculated energy levels near the threshold, I obtain several excited states 
having significant component of a spatially developed $\alpha$ cluster around the $^{10}{\rm B}$ cluster. From remarkable $E2$ transitions,  I assign the $^{10}{\rm B}+\alpha$ cluster states to a 
$K^\pi=3^+$ band of 
$J^\pi=3^+$, $4^+$, and $5^+$ states, and a $K^\pi=1^+$ band of 
$J^\pi=1^+$, $2^+$, $3^+$, $4^+$, and $5^+$ states. The former and the latter bands are 
shown by asterisk and down-triangle symbols in Fig.~\ref{fig:spe}. 
The $K^\pi=3^+$ band has the significant $^{10}$B($3^+$)+$\alpha$ component, whereas
the $K^\pi=1^+$ band contains the $^{10}$B($1^+$)+$\alpha$ component. More details of structure of these states are discussed in the next section.

Figure \ref{fig:be2} shows $E2$ transitions with 
$B(E2)\ge 15$ ${\rm e}^2{\rm fm}^4$ for $J \rightarrow J-1$ and 
$J\rightarrow J-2$ transitions. In-band transitions for the $K^\pi=3^+$ and $^\pi=1^+$ 
$^{10}{\rm B}+\alpha$ bands
are rather strong because of the developed cluster structures, though $E2$ strengths are 
somewhat fragmented into neighboring states. 

\section{Discussion}\label{sec:discussion}
$^{10}$B+$\alpha$ cluster states in the $K^\pi=3^+$ and $K^\pi=1^+$ bands 
have maximum amplitudes of $\alpha$-cluster probability around $D_\alpha=5$ fm.
In this section, I focus on angular motion of the $\alpha$ cluster
at $D_\alpha=5$ fm. I first investigate the angular momentum coupling of the $\alpha$-cluster ($L_\alpha$)
and the $^{10}$B cluster ($I$) in a weak coupling picture and estimate $\alpha$-decay widths. 
Then, I discuss geometric configurations of $^{10}$B+$\alpha$ cluster states in the strong coupling picture
by analyzing $\theta_\alpha$-dependence of the $\alpha$-cluster probability 
around the deformed $^{10}$B cluster.

\subsection{$D_\alpha$-fixed calculation} 
\begin{figure}[htb]
\begin{center}
\includegraphics[width=6.0cm]{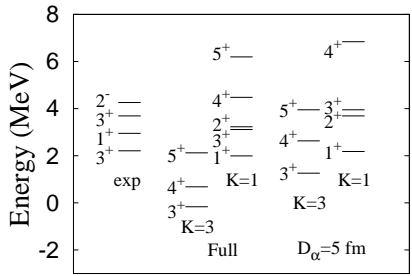} 	
\end{center}
%\vspace{0.5cm}
  \caption{Energies of $^{10}$B+$\alpha$ cluster states 
obtained by the full and $D_\alpha$-fixed calculations and those
observed by the experiment of $^{10}{\rm B}(\alpha,\alpha)^{10}{\rm B}$ reactions \cite{Mo:1973zzb}. 
Energies are measured from the $\alpha$-decay threshold.  
\label{fig:spe-d5}}
\end{figure}

In the present calculation, radial motion of the $\alpha$ cluster is described by
superposing $^{10}$B+$\alpha$ wave functions for $D_\alpha=2,\ldots,6$ fm.
Instead of the full model space in Eq.~\eqref{eq:14N} including $D_\alpha=2,\ldots,6$ fm wave functions,
I also perform a similar calculation using the $D_\alpha$-fixed model space
\begin{equation}\label{eq:14N-D5}
\Psi^{D_\alpha=5}_{^{14}\textrm{N}(J^\pi_n)}=\sum_K \sum_{\theta_\alpha} \sum_{d,\phi} 
C(K,\theta_\alpha,d,\phi) 
\hat P^{J\pi}_{MK} \Phi_{^{10}{\rm B}+\alpha}(D_\alpha,\theta_\alpha; d,\phi),
\end{equation}
where I fix $D_\alpha=5$ fm and 
take $\theta_\alpha=\{0, \pi/8,\pi/4,3\pi/8,\pi/2\}$,  $d=\{1,2\}$ (fm), and
$\phi=\frac{\pi}{4}(j-0.5)$ ($j=1,\ldots,8$). 
In the $D_\alpha=5$ fm fixed calculation, I find the states near the threshold energy corresponding to 
$^{10}$B+$\alpha$ cluster states in the $K^\pi=3^+$ and $K^\pi=1^+$  bands,
but do not obtain lower states below
the threshold because of the truncation of the model space.
Energy levels of the $K^\pi=3^+$ and $K^\pi=1^+$  bands obtained with the full and $D_\alpha$-fixed 
calculations are shown in Fig.~\ref{fig:spe-d5}.
The calculated energies are measured from the $\alpha$-decay threshold.  The experimental
levels observed by $\alpha$ elastic scattering by $^{10}$B are also shown in the figure.
The level structures of the $K^\pi=3^+$ and $K^\pi=1^+$ bands are essentially consistent between 
the full and $D_\alpha$-fixed  calculations, though about 2 MeV global shift is found for 
the $K^\pi=3^+$ band between two calculations. 

%%%%%%%%%%%%%%%%%%%%%%%%%%%%%
%%%%%%%%%%%%%%%%%%%%%%%%%%%%%

\subsection{$\alpha$-cluster probability and $\alpha$-decay widths}
In Table \ref{tab:s-fac}, I show $L_\alpha$ components 
($P_{^{10}{\rm B}(I^\pi)\otimes L_\alpha}$  in Eq.~\eqref{eq:Lalpha}) at $D_\alpha=5 {\ \rm fm}$ 
coupled with $^{10}{\rm B}(3^+)$ and $^{10}{\rm B}(1^+)$ 
in $^{10}$B+$\alpha$ cluster states obtained by the full and $D_\alpha$-fixed calculations.
In the result of the $D_\alpha$-fixed  calculation, $K^\pi=3^+$ band states 
are dominated by the $^{10}{\rm B}(3^+)\otimes L_\alpha$ component, 
whereas $K^\pi=1^+$ band states contain dominantly the
$^{10}{\rm B}(1^+)\otimes L_\alpha$ component. In the result of the full calculation, 
the $K^\pi=3^+$ and $K^\pi=1^+$ band states still contain significant  
$^{10}{\rm B}(3^+)\otimes L_\alpha$ and  $^{10}{\rm B}(1^+)\otimes L_\alpha$ 
components, respectively, except for the $1^+(K^\pi=1^+)$ state, though
the absolute amplitude of the dominant component decreases because of radial motion and 
state mixing.
The $1^+(K^\pi=1^+)$ state obtained by the full calculation shows a feature quite different
from that obtained by the $D_\alpha$-fixed  calculation. In the $D_\alpha$-fixed calculation, 
the $1^+(K^\pi=1^+)$ state is approximately described by the pure 
$^{10}{\rm B}(1^+)\otimes (L_\alpha=0)$ state, where the orbital angular momentum $(L_\alpha)$ of the $\alpha$
cluster weakly couples to the spin $(I)$ of the $^{10}$B cluster. However, in the full calculation, 
the $1^+(K^\pi=1^+)$ state does not show the weak coupling feature but 
has $^{10}{\rm B}(1^+)\otimes (L_\alpha=0)$, $^{10}{\rm B}(1^+)\otimes (L_\alpha=2)$, and  
$^{10}{\rm B}(3^+)\otimes (L_\alpha=2)$ components with the same order 
showing a strong coupling feature.

\begin{figure}[htb]
\begin{center}
\includegraphics[width=10.5cm]{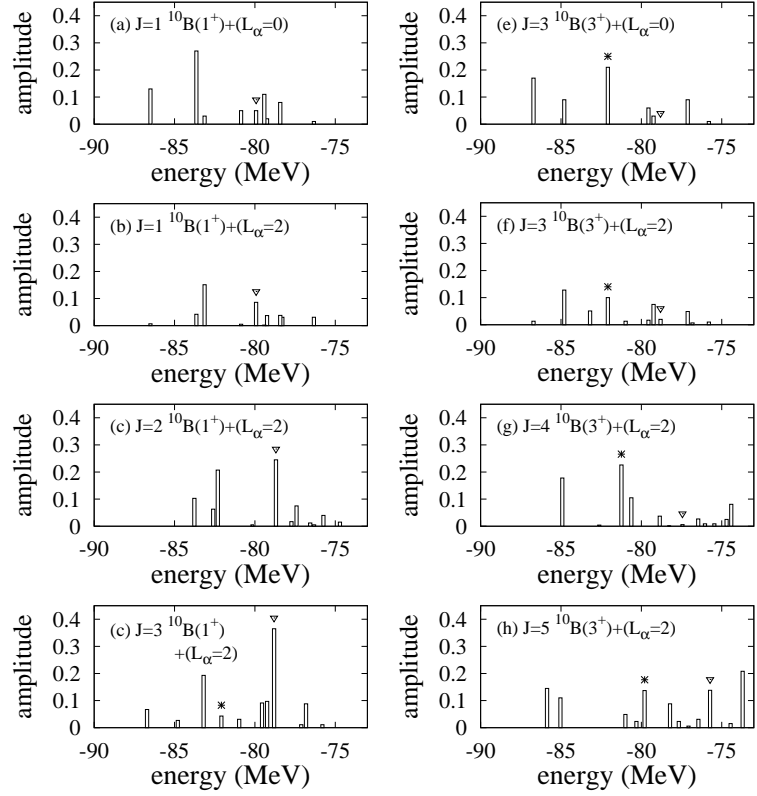} 	
\end{center}
%\vspace{0.5cm}}
  \caption{\label{fig:s-fac-Lalpha}
$^{10}{\rm B}(I^\pi)\otimes (L_\alpha=0,2)$ components, 
$P_{^{10}{\rm B}(I^\pi)\otimes L_\alpha}(D_\alpha=5 {\rm fm})$, 
in positive-parity states of $^{14}$N obtained by the 
$^{10}$B+$\alpha$-cluster model.
Asterisk and down-triangle symbols show $^{10}$B+$\alpha$ cluster states
 in the $K^\pi=3^+$ and $K^\pi=1^+$ bands, respectively. }
\end{figure}

Figure \ref{fig:s-fac-Lalpha} shows $L_\alpha$ components 
($P_{^{10}{\rm B}(I^\pi)\otimes L_\alpha}$) at $D_\alpha=5$ fm of $J^\pi$ states in the $^{14}$N spectra 
obtained by the full calculation. 
The $^{10}$B$(3^+)$+$(L_\alpha=0)$ and $^{10}$B$(3^+)$+$(L_\alpha=2)$ components concentrate at the $3^+(K^\pi=3^+)$ and 
$4^+(K^\pi=3^+)$ states, respectively, though the components are fragmented into other states. 
The $5^+(K^\pi=3^+)$ state shows rather strong state mixing. 
The $^{10}$B$(1^+)$+$(L_\alpha=2)$ component concentrates at the $2^+(K^\pi=1^+)$ and 
$3^+(K^\pi=1^+)$ states, whereas, the $^{10}$B$(1^+)$+$(L_\alpha=0)$ component feeds lower $1^+$ 
states of $^{14}$N.

%%%%%%%%%%%%%%%%%%%%%%%%%%%%%%
%%%%%%%%%%%%%%%%%%%%%%%%%%%%%

In the experiment of $^{10}{\rm B}(\alpha,\alpha)^{10}{\rm B}$ reactions \cite{Mo:1973zzb}, 
the $3^+$ state at $E_r=1.58$ MeV ($E_x=13.19$ MeV) with the width $\Gamma=0.065$ MeV 
is strongly populated. In the analysis of Ref.~\cite{Mo:1973zzb}, this state is described well by the 
dominant (almost 100\%) $S$-wave $\alpha$-decay 
indicating the significant $^{10}{\rm B}(3^+)\otimes (L_\alpha=0)$ component
of the $3^+$ state.
The $1^+$ state at $E_r=2.11$ MeV ($E_x=13.72$ MeV) is weakly populated
in $^{10}{\rm B}(\alpha,\alpha)^{10}B$ reactions, whereas its $\alpha$-decay into the 
fisrt excited state of $^{10}{\rm B}(1^+)$ was observed
in $^{10}{\rm B}(\alpha,\alpha'\gamma)^{10}B$ reactions \cite{Gallmann69}. These experiments
suggest that the $1^+$ state would contain $^{10}{\rm B}(1^+)\otimes (L_\alpha=0)$ and 
$^{10}{\rm B}(3^+)\otimes (L_\alpha=2)$ components. 

%In the present calculation, we obtain  $^{10}$B+$\alpha$ cluster states near the threshold.
From the experimental $\alpha$-decay properties, 
I tentatively assign the theoretical $3^+(K^\pi=3^+)$  and $1^+(K^\pi=1^+)$ states 
having $^{10}$B+$\alpha$ cluster structures 
to the experimental
$3^+$ ($E^\textrm{exp}_r=1.58$ MeV) and $1^+$ ($E^\textrm{exp}_r=2.11$ MeV) states, 
though the band-head energies $E_r(3^+;K^\pi=3^+)=-0.2$ MeV
and $E_r(1^+;K^\pi=1^+)=2.0$ MeV obtained by the full calculation 
do not necessarily agree to the experimental energies (see Fig.~\ref{fig:spe-d5}). 
I estimate partial $\alpha$-decay widths for $B(I^\pi)\otimes L_\alpha$ channels 
from $P_{^{10}B(I^\pi)\otimes L_\alpha}(D_\alpha=a)$ ($a$ is the channel radius) as follows. 
Using the approximate evaluation of the reduced width amplitude 
proposed in Ref.~\cite{Kanada-En'yo:2014nla}, the reduced width $\gamma^2_\alpha(a)$
is calculated as
\begin{equation}
\gamma^2_\alpha(a)=\frac{\hbar^2}{2\mu a}
\left(\frac{\nu}{2\pi}\frac{A_1A_2}{A_1+A_2}\right)^{1/2}
P_{^{10}B(I^\pi)\otimes L_\alpha}(D_\alpha=a), 
\end{equation}
and the partial $\alpha$-decay width $\Gamma_{^{10}B(I^\pi)+\alpha}$ for $L_\alpha=l$ 
is calculated as 
\begin{eqnarray}
\Gamma_{^{10}B(I^\pi)+\alpha}&=&2P_l(a)\gamma^2_\alpha(a),\\
P_l(a)&=&\frac{ka}{F^2_l(ka)+G^2_l(ka)},
\end{eqnarray}
where  $k=\sqrt{2\mu E}/\hbar$, and 
$F_l$ and $G_l$ are the regular and irregular Coulomb functions, respectively.
Here I use the momentum $k$ of the energy $E=E^\textrm{(adjust)}_r$ 
which is phenomenologically adjusted to the 
experimental energy position because 
it is difficult to quantitatively predict the energy position in the present calculation.
Namely, I adjust the band-head energies of the 
$K^\pi=3^+$ and $K^\pi=1^+$ bands to the experimental energy positions $E^\textrm{exp}_r(3^+)=1.58$ MeV and $E^\textrm{exp}_r(1^+)=2.11$ MeV, 
by a constant shift for each band as 
\begin{eqnarray}
&&E^\textrm{(adjust)}_r(J^+;K^\pi=3^+)= E_r(J^+;K^\pi=3^+)-E_r(3^+;K^\pi=3^+)+E^\textrm{exp}_r(3^+),\\
&&E^\textrm{(adjust)}_r(J^+;K^\pi=1^+)= E_r(J^+;K^\pi=1^+)-E_r(1^+;K^\pi=1^+)+E^\textrm{exp}_r(1^+).
\end{eqnarray}
Calculated partial $\alpha$-decay widths are shown in Table \ref{tab:s-fac}.
I calculate widths for $L_\alpha=0$ and $L_\alpha=2$ channels.
%but not for $L_\alpha \ge 4$ channels
%because decomposition of higher partial waves is not possible by the present $\theta_\alpha$ summation.
$\alpha$-decay widths obtained by the full calculation
are several times smaller than those obtained by the $D_\alpha$-fixed 
calculation because of the suppression of the $\alpha$-cluster probability 
as shown previously. As a result, the $\alpha$-decay width of the $3^+(K^\pi=3^+)$ state 
reduces to be $\Gamma_\alpha=0.05$ MeV with the dominant $^{10}{\rm B}(3^+)\otimes(L_\alpha=0)$ 
decay, which is quantitatively consistent with the experimental 
observation ($\Gamma_\alpha\sim \Gamma=0.065(10)$ MeV)  \cite{Mo:1973zzb}. 
For the $1^+(K^\pi=1^+)$ state, I obtain a small $\alpha$-decay width $\Gamma_\alpha=0.01$ MeV with the dominant 
$^{10}{\rm B}(1^+)\otimes(L_\alpha=0)$ decay.
This result seems consistent with the weak population in the $\alpha$ elastic scattering \cite{Mo:1973zzb}
and the fact that the $1^+$ state was observed in 
$^{10}{\rm B}(\alpha,\alpha'\gamma)^{10}B$ reaction \cite{Gallmann69}. However, 
experimental information of partial $\alpha$-decay widths is not enough to confirm the present assignment of
the $1^+(K^\pi=1^+)$ state. The calculated $\alpha$-decay width is much smaller than 
the experimental total width, $\Gamma=0.16(2)$ MeV, of the $1^+$ state at $2.11$ MeV.
I should comment that, because the $^{10}{\rm B}(1^+)\otimes(L_\alpha=0)$ component is 
fragmented into neighboring states as shown in Fig.~\ref{fig:s-fac-Lalpha}, an effectively large 
width could be observed for the $1^+(K^\pi=1^+)$ state.

\begin{table}[htb]
\caption{
\label{tab:s-fac} 
$^{10}{\rm B}(I^\pi)\otimes (L_\alpha=0,2)$ components, 
$P_{^{10}{\rm B}(I^\pi)\otimes L_\alpha}(D_\alpha=5 {\ \rm fm})$,
of  $^{10}$B+$\alpha$ cluster states in the $K^\pi=3^+$ and $K^\pi=1^+$ bands
obtained by the full and $D_\alpha$-fixed calculations. }
\begin{center}
\begin{tabular}{ccccc}
\hline
 & \multicolumn{2}{c}{$P_{^{10}{\rm B}(3^+)\otimes L_\alpha}$}  
&\multicolumn{2}{c}{$P_{^{10}{\rm B}(1^+)\otimes L_\alpha}$}  \\
$J^\pi$ &	$L_\alpha=0$	&	$L_\alpha=2$	&	$L_\alpha=0$	&	$L_\alpha=2$	 \\
 \multicolumn{5}{l} {full cal.}  \\
$3^+$($K^\pi=3^+$)	&	0.21 	&	0.10 	&		&	0.04 			\\
$4^+$($K^\pi=3^+$)	&		&	0.23 	&		&				\\
$5^+$($K^\pi=3^+$)	&		&	0.14 	&		&				\\
$1^+$($K^\pi=1^+$)	&		&	0.03 	&	0.05 	&	0.09 			\\
$2^+$($K^\pi=1^+$)	&		&	0.02 	&		&	0.25 			\\
$3^+$($K^\pi=1^+$)	&	0.00 	&	0.02 	&		&	0.37 			\\
$4^+$($K^\pi=1^+$)	&		&	0.01 	&		&				\\
$5^+$($K^\pi=1^+$)	&		&	0.14 	&		&				\\
 \multicolumn{5}{l} {$D_\alpha$-fixed  cal.}  \\
$3^+$($K^\pi=3^+$)		&	0.57 	&	0.25 	&		&	0.01 			\\
$4^+$($K^\pi=3^+$)		&		&	0.73 	&		&				\\
$5^+$($K^\pi=3^+$)		&		&	0.75 	&		&				\\
$1^+$($K^\pi=1^+$)		&		&	0.02 	&	0.89 	&	0.05 			\\
$2^+$($K^\pi=1^+$)		&		&	0.01 	&		&	0.78 			\\
$3^+$($K^\pi=1^+$)		&	0.10 	&	0.13 	&		&	0.74 			\\
$4^+$($K^\pi=1^+$)		&		&	0.00 	&		&				\\
\hline	
\end{tabular}
\end{center}
\end{table}

\begin{table}[htb]
\caption{
\label{tab:gamma} 
Partial $\alpha$-decay widths 
of $^{10}$B+$\alpha$ cluster states in the $K^\pi=3^+$ and $K^\pi=1^+$ bands
obtained by the full and $D_\alpha$-fixed calculations.
Energies of the band-head states of the $K^\pi=3^+$ and 
$K^\pi=1^+$ bands are adjusted to the experimental resonance energies 
of the $3^+$ state at 1.58 MeV and the $1^+$ state at 2.11 MeV. 
The sum  ($\Gamma_{^{10}{\rm B}+\alpha}(L_\alpha\le 2)$) of  
partial widths of decay channels $^{10}{\rm B}(3^+)\otimes (L_\alpha\le 2)$ and
$^{10}{\rm B}(1^+)\otimes (L_\alpha\le 2)$ is also shown.
The unit is MeV.
}
\begin{center}
\begin{tabular}{ccccccc}
\hline
 & & \multicolumn{2}{c}{$\Gamma_{^{10}{\rm B}(3^+)+\alpha}$} &\multicolumn{2}{c}{$\Gamma_{^{10}{\rm B}(1^+)+\alpha}$} & $\Gamma_{^{10}{\rm B}(3^+)+\alpha}(L_\alpha\le 2)$ \\
$J^\pi$ & $E^{\rm (adjust)}_r$	&$L_\alpha=0$	&	$L_\alpha=2$	&	$L_\alpha=0$	&	$L_\alpha=2$ &  \\
 \multicolumn{5}{l}{full cal.} & \\
$3^+$($K^\pi=3^+$)	&	1.58 	&	0.04 	&	0.00 	&		&	0.00 	&	0.05 	\\
$4^+$($K^\pi=3^+$)	&	2.43 	&		&	0.06 	&		&		&	0.06 	\\
$5^+$($K^\pi=3^+$)	&	3.87 	&		&	0.16 	&		&		&	0.16 	\\
$1^+$($K^\pi=1^+$)	&	2.11 	&		&	0.00 	&	0.01 	&	0.00 	&	0.01 	\\
$2^+$($K^\pi=1^+$)	&	3.35 	&		&	0.02 	&		&	0.09 	&	0.11 	\\
$3^+$($K^\pi=1^+$)	&	3.23 	&	0.00 	&	0.01 	&		&	0.12 	&	0.13 	\\
$4^+$($K^\pi=1^+$)	&	4.60 	&		&	0.01 	&		&		&	0.01 	\\
$5^+$($K^\pi=1^+$)	&	6.31 	&		&	0.36 	&		&		&	0.36 	\\
\multicolumn{5}{l} {$D_\alpha$-fixed  cal.}&  \\
$3^+$($K^\pi=3^+$)	&	1.58 	&	0.12 	&	0.01 	&		&	0.00 	&	0.13 	\\
$4^+$($K^\pi=3^+$)	&	2.95 	&		&	0.41 	&		&		&	0.41 	\\
$5^+$($K^\pi=3^+$)	&	4.27 	&		&	1.07 	&		&		&	1.07 	\\
$1^+$($K^\pi=1^+$)	&	2.11 	&		&	0.00 	&	0.10 	&	0.00 	&	0.11 	\\
$2^+$($K^\pi=1^+$)	&	3.61 	&		&	0.01 	&		&	0.41 	&	0.42 	\\
$3^+$($K^\pi=1^+$)	&	3.88 	&	0.19 	&	0.15 	&		&	0.51 	&	0.85 	\\
$4^+$($K^\pi=1^+$)	&	6.77 	&		&	0.01 	&		&		&	0.01 	\\
\hline	
\end{tabular}
\end{center}
\end{table}

\subsection{Angular motion of the $\alpha$ cluster around the deformed $^{10}$B cluster}

I here discuss angular motion of 
the $\alpha$-cluster around the deformed $^{10}$B cluster
by analyzing $\theta_\alpha$ dependence of $\alpha$-cluster probabilities. 
Discussions in this section are based on the strong coupling picture, which is somehow different
from the previous discussion based on the 
$L_\alpha$ decomposition in the weak coupling picture.
I show energies of 
$\Phi_{^{10}{\rm B}(I^\pi_z)+\alpha}(D_\alpha,\theta_a)$, in which the $\alpha$ cluster
is localized at $(D_\alpha,\theta_\alpha)$ around the $I_z$ projected $^{10}$B cluster. 
In Fig.~\ref{fig:s-fac-Lalpha}, intrinsic energies before parity and angular-momentum projections of 
$\Phi_{^{10}{\rm B}(I^\pi_z)+\alpha}(D_\alpha,\theta_a)$ 
for $I^\pi_z=3^+$ and $1^+$ 
are plotted on the $(x,z)=(D_\alpha \sin\theta_\alpha,D_\alpha \cos\theta_\alpha)$ plane.
The energy curves for $D_\alpha=5$ fm are also shown as functions of $\theta_\alpha$. 
In the $D_\alpha \ge 5$ fm region, the contour of the energy surface on the $(x,z)$ plane 
is deformed in the longitudinal ($\theta_\alpha=0$) 
direction because of the prolate deformation of the $^{10}$B cluster meaning that
the $\alpha$ cluster at the fixed distance $D_\alpha=5$ fm feels an attraction in the longitudinal direction. 
In other words, in the intrinsic system, the $\alpha$ cluster at $D_\alpha=5$ fm energetically 
favors the longitudinal direction to form the linear $3\alpha$ configuration rather than the transverse direction 
to form the triangle $3\alpha$ configuration. 
In the $D_\alpha \le 3$ fm region, the $\alpha$ cluster feels an effective repulsion in the longitudinal direction because of the Pauli blocking from the $^{10}$B cluster, whereas it feels an attraction
in the transverse ($\theta_\alpha=\pi/2$) direction.

\begin{figure}[htb]
\begin{center}
\includegraphics[width=8.0cm]{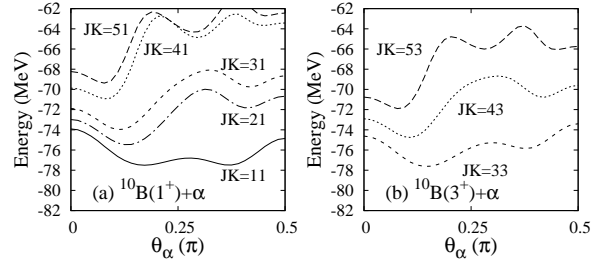} 	
\end{center}
 \caption{\label{fig:ene-theta}
Energies of the $JK$-projected $\Phi_{^{10}{\rm B}(I^\pi_z)+\alpha}$ wave function
$\hat P^{J\pi}_{MK}\Phi_{^{10}{\rm B}(I^\pi_z)+\alpha}(D_\alpha,\theta_a)$ with 
$K=I_z$ for (a) $^{10}{\rm B}(I^\pi_z=3^+)$ and (b) $^{10}{\rm B}(I^\pi_z=1^+)$. 
Energies for $D_\alpha=5$ fm are plotted as functions of $\theta_\alpha$. 
}
\end{figure}

In contrast to the intrinsic energy behavior, 
$\theta_\alpha$ dependence of $J^\pi$-projected energy is not trivial
because the energy is affected by not only  potential energy but also 
kinetic energy of angular motion, i.e.,  rotational energy.
Figure \ref{fig:ene-theta} shows energies of $JK$-projected states 
$\hat P^{J\pi}_{MK}\Phi_{^{10}{\rm B}(I_z)+\alpha}(D_\alpha,\theta_a)$ at $D_\alpha=5$ fm
for $K=I_z$, which corresponds to the $L_{\alpha z}=0$ projection. 
In high $J$ states, the longitudinal direction ($|\theta_\alpha| \lesssim \pi/8$) is energetically favored 
than the transverse direction ($|\theta_\alpha-\pi/2| \lesssim \pi/8$)
because of the 
larger moment of inertia (m.o.i.) of the longitudinal configuration than 
that of the transverse configuration for the $L_{\alpha z}=0$ projection.
However, in the lowest spin state ($JK=11$), the energy almost degenerates
in a wide region of $\theta_\alpha$ because the kinetic energy is smaller for the transverse configuration 
than the longitudinal configuration because of the phase space factor $\sin \theta_\alpha$ in the
$L_{\alpha z}=0$ projection. This energy degeneracy results in the 
$L_\alpha=0$ ($S$-wave) dominance in the $1^+(K^\pi=1^+)$ state obtained by the $D_\alpha$-fixed calculation.

\begin{figure}[htb]
\begin{center}
\includegraphics[width=12.0cm]{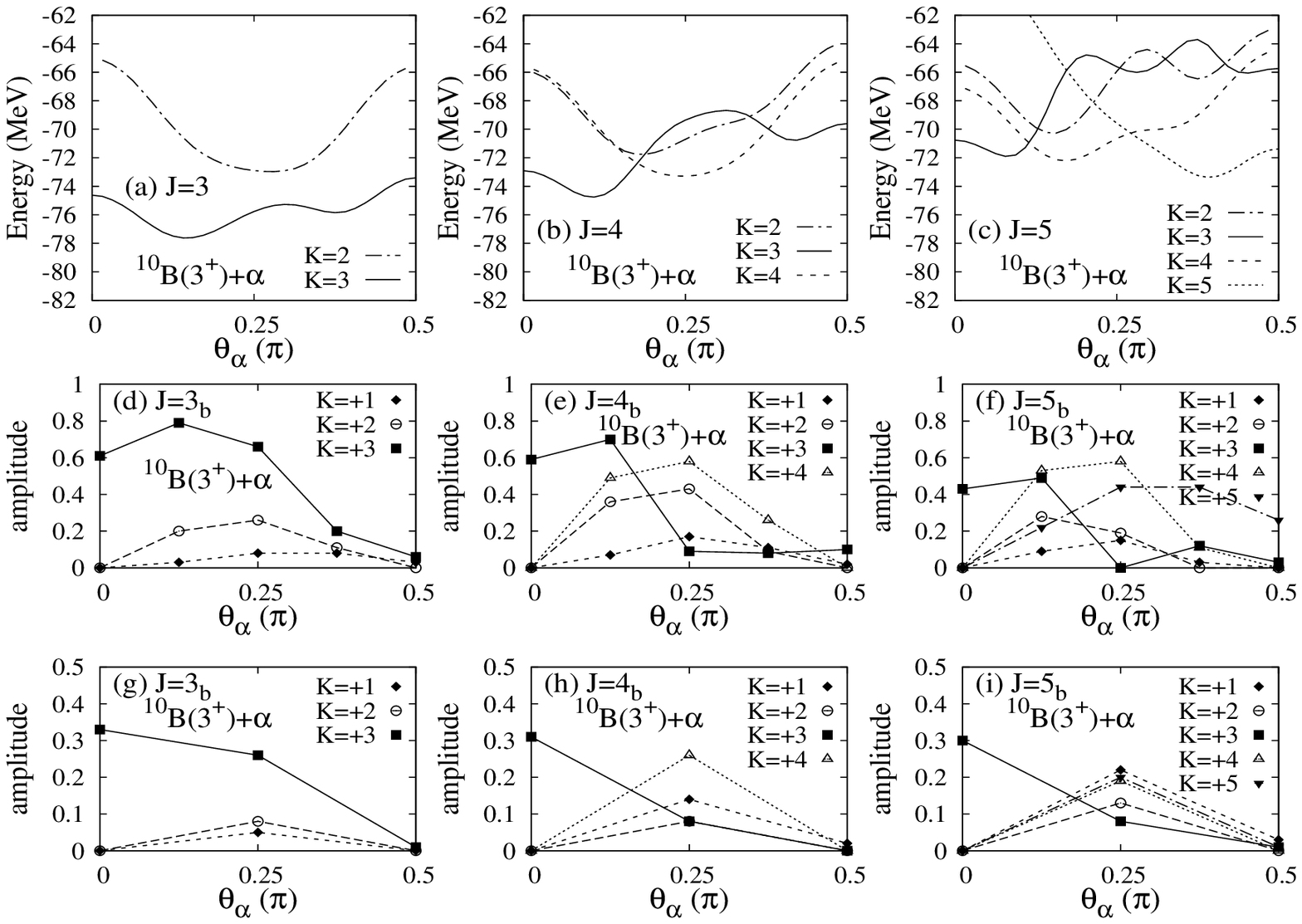} 	
\end{center}
\caption{\label{fig:ene-theta-J3}
(a)(b)(c) Energies of the $JK$-projected $\Phi_{^{10}{\rm B}(I^\pi_z)+\alpha}$ wave function
$\hat P^{J\pi}_{MK}\Phi_{^{10}{\rm B}(I^\pi_z)+\alpha}(D_\alpha,\theta_a)$ for $^{10}{\rm B}(I^\pi_z=3^+)$. 
(d)(e)(f) $\alpha$-cluster probability 
$P(JK; ^{10}{\rm B}(I^\pi_z); D_\alpha, \theta_\alpha)$ for $I^\pi_z=3^+$ at 
$D_\alpha=5$ fm in $K^\pi=3^+$ $^{10}$B+$\alpha$ cluster states 
obtained by the $D_\alpha$-fixed calculation and (g)(h)(i) that obtained by the full calculation. 
}
\end{figure}

\begin{figure}[htb]
\begin{center}
\includegraphics[width=16.0cm]{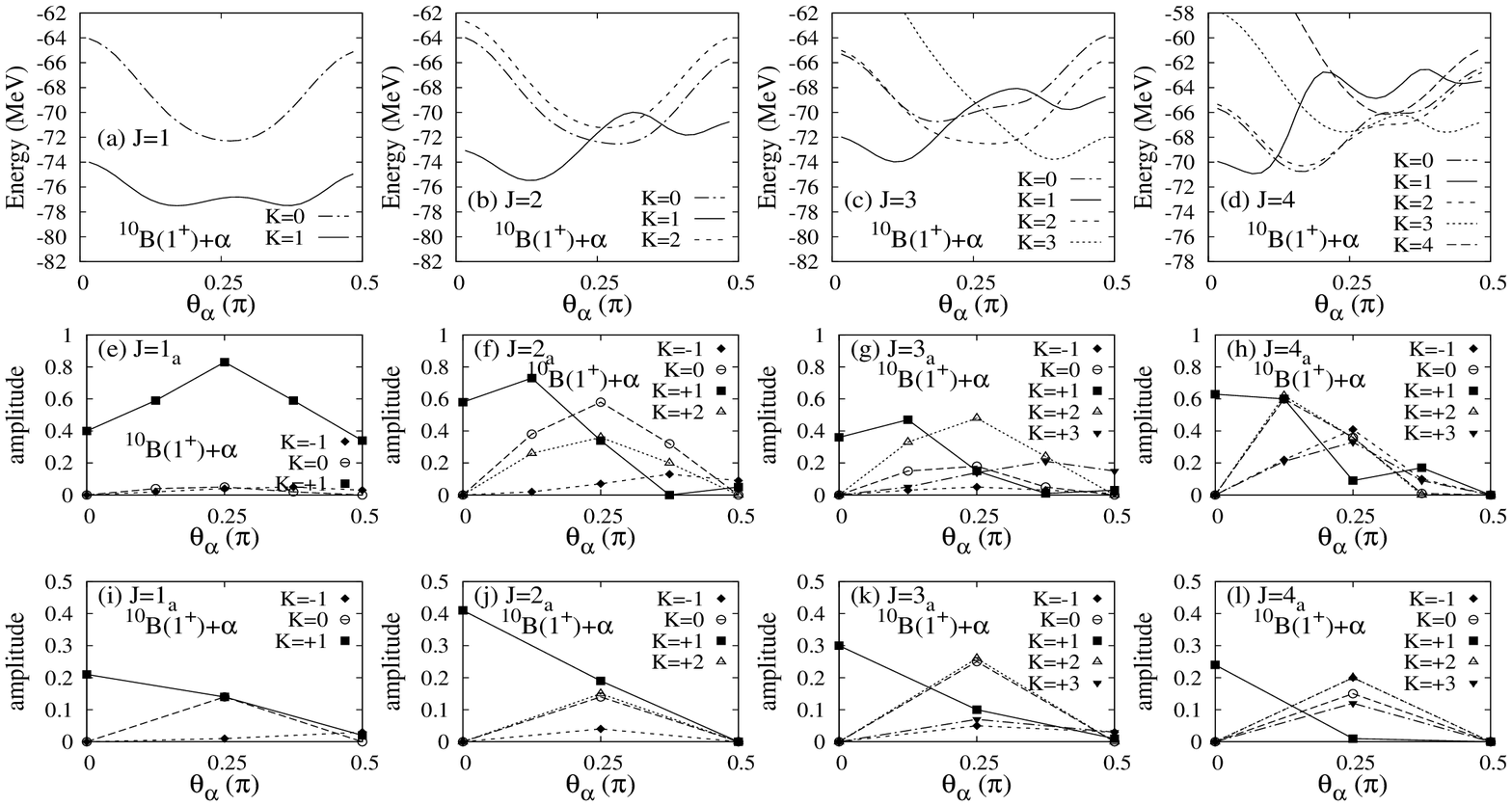} 	
\end{center}
 \caption{\label{fig:ene-theta-J1}
(a)-(d) Energies of the $JK$-projected $\Phi_{^{10}{\rm B}(I^\pi_z)+\alpha}$ wave function
$\hat P^{J\pi}_{MK}\Phi_{^{10}{\rm B}(I^\pi_z)+\alpha}(D_\alpha,\theta_a)$ for $^{10}{\rm B}(I^\pi_z=1^+)$. 
(e)-(h) $\alpha$-cluster probability 
$P(JK; ^{10}{\rm B}(I^\pi_z); D_\alpha, \theta_\alpha)$ for $I^\pi_z=1^+$ at 
$D_\alpha=5$ fm in the $K^\pi=1^+$ $^{10}$B+$\alpha$ cluster states 
obtained by the $D_\alpha$-fixed calculation and (i)-(l) that obtained by the full calculation. 
}
%\vspace{0.5cm}}
\end{figure}

Figures ~\ref{fig:ene-theta-J3} and \ref{fig:ene-theta-J1} show energies of $JK$-projected states 
at $D_\alpha=5$ fm for $K\ne I_z$.  Note that
the $K\ne I_z$ projection corresponds to the $L_{\alpha z}\ne 0$ projection, 
and $K> I_z$ means the $L_{\alpha}$ alignment to the $z$ direction (see Fig.~\ref{fig:10B-4He}(c)). 
For instance, the $L_\alpha$-aligned state for $L_\alpha=2$ ($D$-wave) is the $K=I_z+2$ state.
As shown in Figs.~\ref{fig:ene-theta-J3}(a)-(c) and \ref{fig:ene-theta-J1}(a)-(d), 
$L_\alpha$-aligned states energetically favor the transverse configuration because of the larger m.o.i. 
than that of the longitudinal configuration in the $L_{\alpha z}=2$ projection.
Figures ~\ref{fig:ene-theta-J3} and \ref{fig:ene-theta-J1} also show the 
$\alpha$-cluster probability $P(JK; ^{10}{\rm B}(I^\pi_z); D_\alpha, \theta_\alpha)$ at 
$D_\alpha=5$ fm in the $^{10}$B+$\alpha$ cluster states obtained by the $D_\alpha$-fixed and full calculations. 
Let me first discuss the result obtained by the $D_\alpha$-fixed calculation (Figs.~\ref{fig:ene-theta-J3}(d)-(f)
and \ref{fig:ene-theta-J1}(e)-(h)). 
In the $K^\pi=3^+$ band states (Fig.~\ref{fig:ene-theta-J3}(d)-(f)), 
the $J^\pi=3^+$ state contains dominantly the longitudinal configuration 
($|\theta_\alpha| \lesssim \pi/8$) rather than the transverse configuration ($|\theta_\alpha-\pi/2| \lesssim \pi/8$)
as expected from the $JK$-projected energy curve for $K=I_z$. As $J$ goes up to $J=5$, the $L_\alpha$-aligned 
component $(K=5)$ of the transverse configuration becomes large corresponding to the alignment of the orbital 
angular momentum $L_\alpha$ of the $\alpha$ cluster to $I_z=3$ (the spin of $(pn)$ cluster in the $^{10}$B cluster).
In the $K^\pi=1^+$ band states (Fig.~\ref{fig:ene-theta-J1}(e)-(h)),
the $J^\pi=1^+$ state shows the $\alpha$-cluster
probability distributed widely in the $0 \le \theta_\alpha \le \pi/2$ region indicating the dominant $L_\alpha=0$ ($S$-wave) component.
As $J$ increases, the longitudinal component becomes dominant compared with
the transverse component. The alignment of $L_\alpha$ (the orbital angular momentum of the $\alpha$ cluster) 
and $I_z$ is not so remarkable for $^{10}$B$(I^\pi_z=1^+)$ differently from the
$^{10}$B$(I^\pi_z=3^+)$. 
Next, let me look into the result of the full calculation shown in Figs.~\ref{fig:ene-theta-J3}(g)-(i) and \ref{fig:ene-theta-J1}(i)-(l). 
Compared with the $D_\alpha$-fixed  calculation, transverse components tend to be relatively more suppressed 
than longitudinal components in the full calculation. 
%As a result, the longitudinal
%component becomes larger than the transverse component. 
Note that the longitudinal ($\theta_\alpha=0$) component is not dominant 
but is $30\sim 40\%$, which is comparable to the $\theta_\alpha=\pi/4$ component.
It indicates that $^{10}$B+$\alpha$ cluster states are different from the ideal linear configuration
of a classical picture but they show significant quantum fluctuation in the angular ($\theta_\alpha$) motion
and are regarded as the chain-like configuration that has relatively enhanced longitudinal components 
with suppressed transverse components.

%%%%%%%%%%%%%%%%%%%%%%%%%%%%%%
\begin{figure}[htb]
\begin{center}
\includegraphics[width=5.5cm]{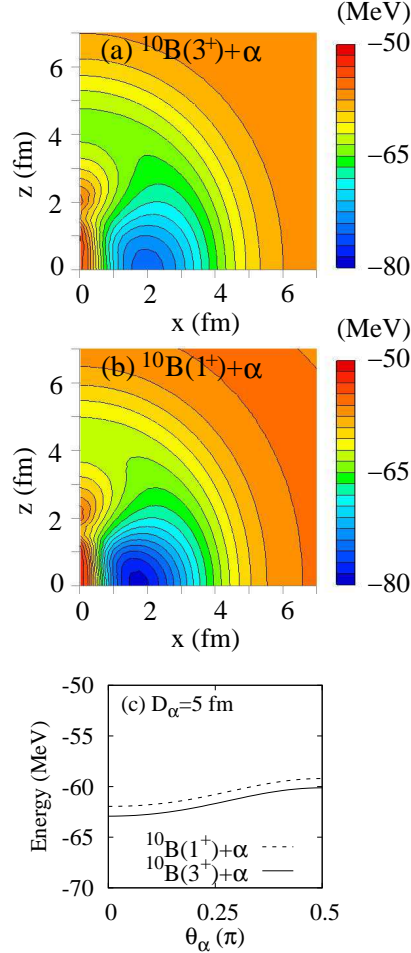} 	
\end{center}
 \caption{\label{fig:ene-nopro}
Intrinsic energies of $^{10}{\rm B}(I^\pi_z=3^+)+\alpha$ and
$^{10}{\rm B}(I^\pi_z=1^+)+\alpha$ before the parity and angular-monentum projections.
Energies for (a) $^{10}{\rm B}(I^\pi_z=3^+)+\alpha$ and (b) $^{10}{\rm B}(I^\pi_z=1^+)+\alpha$ plotted 
on $(x,z)=(D_\alpha \sin\theta_\alpha,D_\alpha \cos\theta_\alpha)$,
and (c) those at $D_\alpha=5$ fm plotted as functions of $\theta_\alpha$. 
}
%\vspace{0.5cm}}
\end{figure}
%%%%%%%%%%%%%%%%%%%%%%%%%%%%%

%%%%%%%%%%%%%%%%%%%%%%%%%%%%%%
%%%%%%%%%%%%%%%%%%%%%%%%%%%%%

%%%%%%%%%%%%%%%%%%%%%%%%%%%%%%
%%%%%%%%%%%%%%%%%%%%%%%%%%%%%

%%%%%%%%%%%%%%%%%%%%%%%%%%%%%%
%%%%%%%%%%%%%%%%%%%%%%%%%%%%%

\begin{figure}[htb]
\begin{center}
\includegraphics[width=10.5cm]{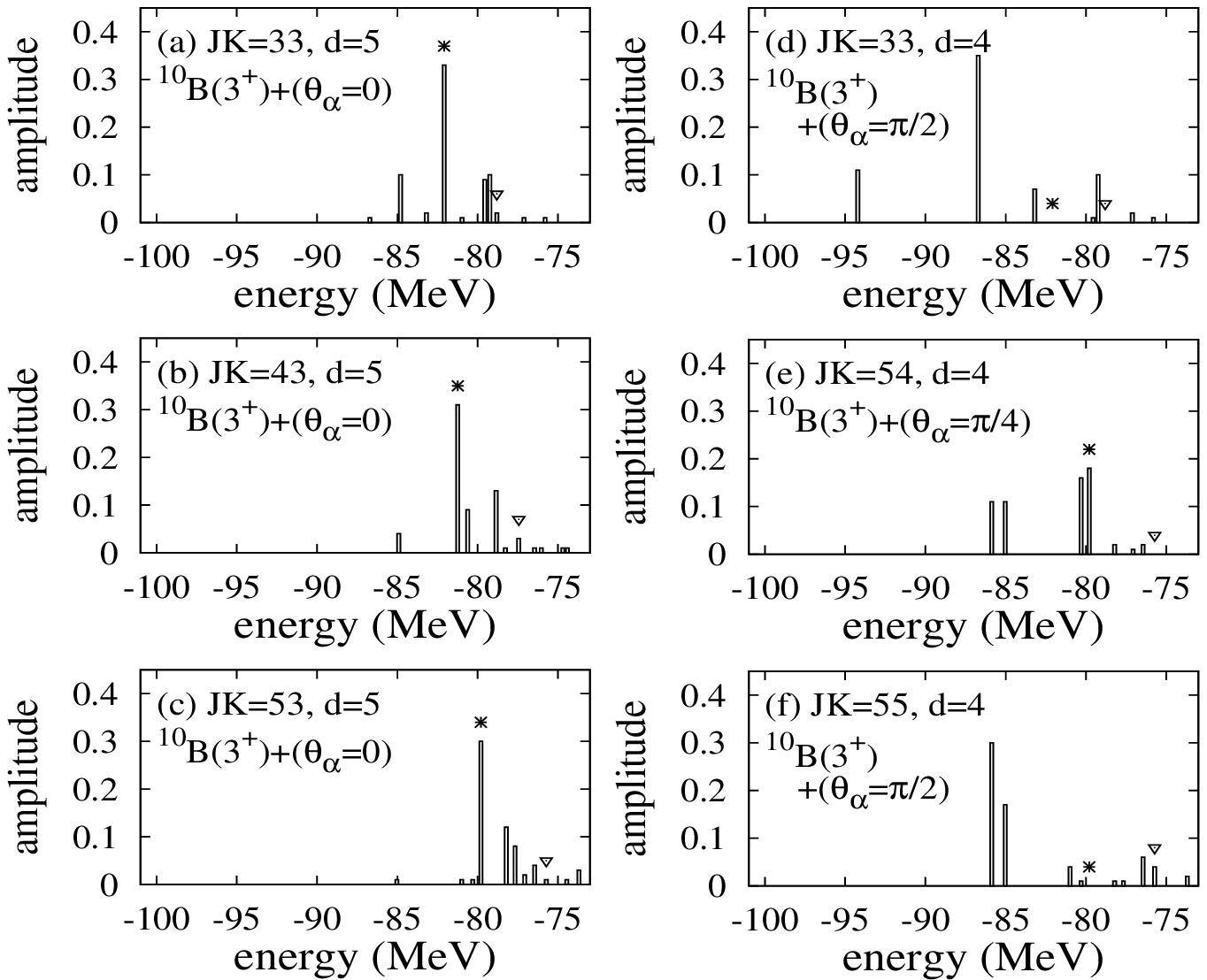} 	
\end{center}
\caption{\label{fig:s-theta-J3}
$\alpha$-cluster probability $P(JK; ^{10}{\rm B}(I^\pi_z); D_\alpha, \theta_a)$ 
for $I^\pi_z=3^+$. $D_\alpha$ is taken to be $D_\alpha=5$ fm for $\theta_\alpha=0$, and 
$D_\alpha=4$ fm for $\theta_\alpha=\pi/4$ and $\pi/2$.
Asterisk and down-triangle symbols show $^{10}$B+$\alpha$ cluster states
in the $K^\pi=3^+$ and $K^\pi=1^+$ bands, respectively. 
}
%\vspace{0.5cm}}
\end{figure}

\begin{figure}[htb]
\begin{center}
\includegraphics[width=10.5cm]{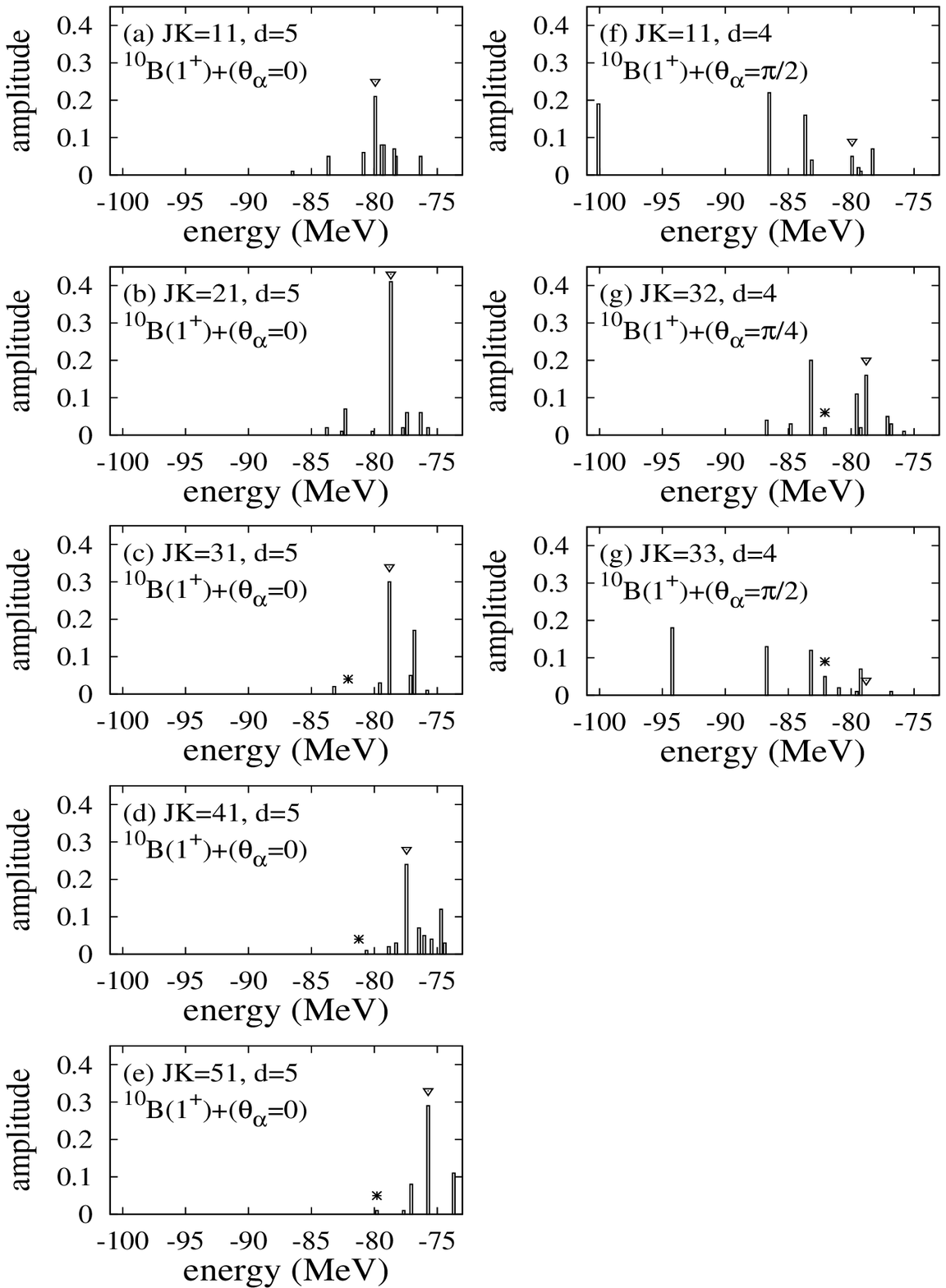} 	
\end{center}
\caption{\label{fig:s-theta-J1}
$\alpha$-cluster probability $P(JK; ^{10}{\rm B}(I^\pi_z; D_\alpha, \theta_a)$ 
for $I^\pi_z=1^+$. $D_\alpha$ is taken to be $D_\alpha=5$ fm for $\theta_\alpha=0$, and 
$D_\alpha=4$ fm for $\theta_\alpha=\pi/4$ and $\pi/2$.
Asterisk and down-triangle symbols show $^{10}$B+$\alpha$ cluster states
 in the $K^\pi=3^+$ band $K^\pi=1^+$ bands, respectively. 
}
\end{figure}

The origin of the suppression of transverse components in $^{10}$B+$\alpha$ cluster 
states in the full calculation can be described by orthogonality to lower states which 
contain transverse components with $D_\alpha < 5$ fm. 
As shown in Fig.~\ref{fig:ene-nopro} for the energy surface on the $(D_\alpha,\theta_a)$ plane, 
an energy pocket exists in the transverse direction ($\theta_\alpha\sim \pi/2$) around $D_\alpha\sim 2$,
and therefore, transverse components contribute to low-lying $^{14}$N states. 
Although the low-lying states are compact states containing  mainly configurations with 
small $D_\alpha$, transverse components with $D_\alpha=5$ fm 
somewhat feed the low-lying states. 
As a result of the feeding of lower states, transverse components in the $^{10}$B+$\alpha$ cluster states
near the threshold are suppressed. 
Figures \ref{fig:s-theta-J3} and \ref{fig:s-theta-J1} 
show the $\alpha$-cluster probability $P(JK; ^{10}{\rm B}(I^\pi_z); D_\alpha, \theta_a)$ for $\theta_\alpha=0$ 
at $D_\alpha=5$ fm and that for $\theta_\alpha=\pi/4$ and $\pi/2$  at $D_\alpha=4$ fm.
As seen in \ref{fig:s-theta-J3}(a)-(c) for $^{10}{\rm B}(I^\pi_z=3^+)$, the longitudinal ($\theta_\alpha=0$) component 
of  $^{10}{\rm B}(I^\pi_z=3^+)+\alpha$ shows the largest amplitude at the 
$K^\pi=3^+$ band states (labeled by asterisks) and some fragmentation into neighboring states. Similarly, 
the longitudinal component of  $^{10}{\rm B}(I^\pi_z=1^+)+\alpha$ concentrates on the
$K^\pi=1^+$ band states (see Fig.~\ref{fig:s-theta-J1}(a)-(e)). 
On the other hand,  transverse components feed states lower than 
$^{10}$B+$\alpha$-cluster states as seen in Fig.~\ref{fig:s-theta-J3}(d)(f) and Fig.~\ref{fig:s-theta-J1}(f)(g)).
Consequently the $\alpha$ cluster in 
$^{10}$B+$\alpha$-cluster states near the threshold tends to avoid transverse configurations so as
to satisfy orthogonality to lower states. This mechanism is 
consistent with the discussion of Ref.~\cite{Suhara:2010ww} for linear-chain 3$\alpha$ states in $^{14}$C. 

%%%%%%%%%%%%%%%%%%%%%%%%%%%%%%
%%%%%%%%%%%%%%%%%%%%%%%%%%%%%

%%%%%%%%%%%%%%%%%%%%%%%%%%%%%%
%%%%%%%%%%%%%%%%%%%%%%%%%%%%%

\section{Summary} \label{sec:summary}
I calculated positive-parity states of $^{14}$N with the $^{10}$B+$\alpha$ 
cluster model and investigated $^{10}$B+$\alpha$ cluster states. 
Near the $\alpha$-decay threshold energy, I obtained the 
$K^\pi=3^+$ and $K^\pi=1^+$ rotational bands
having the developed $\alpha$ cluster with the $^{10}$B($3^+$) and $^{10}$B($1^+$) cores, 
respectively.
I assigned the $3^+(K^\pi=3^+)$ state in the present result to 
the experimental $3^+$ at $E_r=1.58$ MeV 
observed in $\alpha$ scattering reactions by $^{10}$B, and showed that the calculated 
$\alpha$-decay width agrees to the experimental width.

I analyzed the component of the longitudinal configuration having
an $\alpha$ cluster in the longitudinal direction of the deformed $^{10}$B cluster, which 
corresponds to a linear-chain $3\alpha$ structure with valence nucleons. 
In the spectra of $^{14}$N, the linear-chain component concentrates at the 
$^{10}$B+$\alpha$ cluster states in the $K^\pi=3^+$ and $K^\pi=1^+$ bands.
However, the $^{10}$B+$\alpha$ cluster states are different from the ideal linear configuration
of a classical picture but they show significant quantum fluctuation in the angular ($\theta_\alpha$) motion
and are regarded as the chain-like configuration that has relatively enhanced longitudinal components
and suppressed transverse components.
The orthogonality to low-lying states plays an essential role in the suppression of the
transverse component. 

\section*{Acknowledgments} 
The authors would like to thank Dr.~Suhara for fruitful discussions.
The computational calculations of this work were performed by using the
supercomputers at YITP. This work was supported by 
JSPS KAKENHI Grant Number 26400270.

\end{document}